\documentclass[letterpaper]{article}
\usepackage{natbib,alifeconf}

\usepackage{subfigure}
\usepackage{tikz}
\usepackage{tkz-graph}
\usetikzlibrary{calc,%
arrows,plotmarks}
\usepackage{pgfplots}
\usepackage{amsmath, amsthm, amssymb}

\title{Don't Believe Everything You Hear; \\ Preserving Relevant Information by Discarding Social Information}
\author{Christoph Salge$^{1}$ \and Daniel Polani $^1$ \\
\mbox{}\\
$^1$University of Hertfordshire, Hatfield, UK\\
c.salge\textbar d.polani@herts.ac.uk}

\begin{document}
\maketitle

\begin{abstract}
Integrating information gained by observing others via Social Bayesian Learning can be beneficial for an agent's performance, but can also enable population wide information cascades that perpetuate false beliefs through the agent population. We show how agents can influence the observation network by changing their probability of observing others, and demonstrate the existence of a population-wide equilibrium, where the advantages and disadvantages of the Social Bayesian update are balanced. We also use the formalism of relevant information to illustrate how negative information cascades are characterized by processing increasing amounts of non-relevant information. 
\end{abstract}

\section{Introduction}
Information processing is an important aspect of life. Organisms equipped with sensors obtain and utilize information to increase their inclusive fitness; thus justifying the existence of (often costly) sensors in the first place \citep{polani2009information}. 
However, not all information is equally relevant for an organism -- a notion formalised by \citet{Polani2001,Polani2006}, which we will introduce in more detail later. The basic idea of \emph{relevant information} is to quantify how much information at least is needed to obtain a certain performance level. Once this is established, the next question to ask is, how to best obtain this specific information?

Previously, we argued \citep{salgeJASSS} that agents with common goals and embodiments are likely to have similar relevant information. Once they obtain this relevant information, they also have to act upon it to reap its benefits, thus encoding it in their actions. As the state-space of actions is usually much smaller than the state-space of the overall environment, this is likely to lead to a higher ``concentration'' of relevant information in another agent's actions rather than in the environment itself. This \emph{digested information}, encoded in actions, concentrates “pre-processed” decision-relevant information and provides incentives for agents to observe each other and modify their own actions accordingly. However, similar behaviour in a population of agents can lead to a phenomenon called \emph{herding} \citep{banerjee1992simple} or \emph{information cascade} \citep{bikhchandani1992theory}. This usually requires an agent population where agents:
\begin{itemize}
\itemsep0em
\item select one of several choices;
\item have some private information related to their decision;
\item act sequentially and can observe the choices of others, but not the private internal information of others.
\end{itemize}
This can then lead to situations such as the example by \cite{easley2010information}, where an agent wants to choose between restaurant A and B. His own research suggests that restaurant A is better, but once he gets there, no one is eating in restaurant A, while restaurant B is filled with customers. Based on this information it is reasonable to infer that several other agents have private information that caused them to choose B instead of A. By inferring this additional information it becomes rational to choose B instead of A, even if his own private information suggests otherwise. 

The problem here is that others might make similar conclusions, and create a chain reaction of inferred private information that is based on no or very little private information. This illustrates two common properties of information cascades; they can be based on very little initial information, and they can be wrong.


This is somewhat in contrast to the argument presented in ``The Wisdom of Crowds'', where \citet{surowiecki2005wisdom} argues that agents that aggregate their information can produce very accurate results. But, as \citet{easley2010information} point out, this only applies if they are guessing independently. Furthermore, recent studies \citep{kao2014decision} examining several models of group behaviour suggest that small groups make correct decision, while larger groups are more likely to converge on an incorrect decision. Also note that information cascades are also present in other types of multi-agent scenarios, such as swarm coordination \citep{wang2012quantifying}, and are potentially subject to similar problems.

\subsection{Overview}

In this paper we examine the interaction between the positive and negative effects of observing others through the perspective of the relevant information framework. In particular, we show how rational adaptations can lead to a situation where incorrect information cascades become common, and how they are characterized by a reduction in the density of relevant information. Furthermore, we demonstrate that in this environment it is reasonable for agents to randomly discard part of their sensor intake. 

After introducing information theory and relevant information in more detail, we present the single agent model to create a baseline for agent performance and demonstrate how an agent's actions encode information. The multi-agent scenario is then used to motivate the introduction of the Social Bayesian Update, as it demonstrates the increase in performance when information from other agents is used in decision making. The next scenario deals with changing world states and shows that agent's performance can be increased by explicitly modelling the noise in the world, which basically motivates internal models which cannot express certainty. This specific form of bounded rationality is interesting in the context of information cascades, as \citet{acemoglu2011bayesian} previously showed that a lack of internal certainty makes populations more likely to synchronize. 
Finally, we will look at models that combine noise and Social Bayesian Update, which have both been motivated previously by increased agent performance. In this environment, negative information cascades are common but we show that agents can randomly discard sensor inputs to increase their performance. This is motivated by results from \citet{gale2003bayesian}, which demonstrated that sparsity in the observation graph makes convergence (both negative and positive) less likely. By moderating their own sensor intake, agents can change between single-agent behaviour, and positive ``wisdom of the crowds'' and negative information cascades.

\section{Information Theory}

Relevant information is based on the formalism of Information Theory \citep{Shannon1948}. If $X$ is a random variable that can assume the states $x$, where each state $x$ is a member of the alphabet $\mathcal{X}$, then $P(X)$ is the probability distribution of $X$, and $P(X=x)$ is the probability that $X$ assumes the value $x$, sometimes shortened to $p(x)$. Entropy, or the self-information of a variable is then defined as \begin{equation}
H(X)=-\sum_{x \in \mathcal{X}} p(x)\log p(x).
\end{equation}
This is often described as the uncertainty about the outcome of $X$, the average expected surprise, or the average information gained if one was to observe the state of $X$, without having prior knowledge about $X$. Consider two jointly distributed random variables, $X$ and $Y$; then we can calculate the \textit{conditional entropy} of $X$ given a particular outcome $Y=y$ as
\begin{equation}
H(X|Y=y)=-\sum_{x \in \mathcal{X}} p(x|y)\log p(x|y).
\end{equation}
This can be averaged over all states of $Y$, resulting in the conditional entropy of $X$ given $Y$, 
\begin{equation}
H(X|Y)=-\sum_{y \in \mathcal{Y}} p(y) \sum_{x \in \mathcal{X}} p(x|y)\log(p(x|y)).
\end{equation}
This is the entropy of $X$ that remains, on average, if $Y$ is known. So $H(X)$ and $H(X|Y)$ are the entropy of $X$ before and after we learn the state of $Y$. Thus, their difference is the amount of information we can learn, on average, about $X$ by knowing $Y$. Subtracting one from the other, we get a value called \textit{mutual information}: 
\begin{equation}
I(X;Y)=H(X)-H(X|Y).
\end{equation}
The mutual information is symmetrical and measures the amount of information one random variable contains about another (and vice versa, by symmetry). Also, note that we use the binary logarithm for all $\log(.)$ operations, so all information measurements are in \textit{bits}.

\subsection{Relevant Information}

Relevant information is the amount of information an agent needs to obtain to either act optimally, or at a specific performance level. Assume that there is an agent that interacts with the environment by choosing an action in reaction to some form of sensor input. The environment $R$ is in the state $r$, and the agent chooses an action $a$ from a set of actions $A$. For simplicity, we assume for now that the agent can perceive the whole environment, so the sensor state is equal to the state of the environment. 
Furthermore, assume that the actions of the agent are connected to some utility function $U(a,r)$ (for example, survival probability, or fitness) which determines different pay-offs, depending on the agent's action $A=a$ and the state of the environment $R=r$. We also assume that the states of the world $R$ are distributed according to the probability distribution $P(R)$.

A \textit{strategy} is defined as a conditional probability distribution $P(A|R)$, which defines for every state $r$ the probability of choosing different actions $a$. We can define a set $\pi^u$ as the set of all strategies that have the average pay-off level, or performance, of at least $u$ as
\begin{equation}
\pi^u=\left\lbrace P(A|R)\middle|\sum_a \sum_r U(a,r) p(a|r) p(r) \geq u\right\rbrace.
\end{equation} 
As a strategy $P(A|R)$ also implies a distribution $P(A)=P(A|R)P(R)$, we can compute the mutual information $I(A;R)$ for each strategy. The relevant information for a specific performance level is then defined as
\begin{equation}
RI(u) := \min_{p(a|r) \in \pi^u } I(A;R), 
\label{eq:RIU}
\end{equation}
which is the minimal mutual information over all strategies that achieve at least the average pay-off of $u$. As the mutual information $I(A;R)$ measures the amount of information the agent has to process to determine $a$, this can be interpreted as the minimal amount of information an agent needs to obtain to perform at least as well as $u$. Due to the symmetry of mutual information, this can also be interpreted as the minimal amount of information an agent's actions have to contain.

 
\section{Experiments}

\subsection{Single Agent Model}

There are ten locations; exactly one of them contains treasure. The treasure location is modelled by the state of the variable $T$. The agent's task is to determine the location of the treasure in the least number of turns. Each turn the agent decides to visit one of the locations, and is then informed if that location contains the treasure or not.  

The agent's decision making is modelled with an internal Bayesian model $\hat{T}$, where $P(\hat{T}=t)$ is the agent-assumed probability that the treasure is in location $t$. Every turn, the agent chooses to visit the location where it believes the treasure most likely to be. In case of a tie between different locations, it chooses one of them at random. Initially, the agent believes all locations to be equally likely. Once it observes the state of a given location, it updates its internal model with that knowledge. So, if location $t$ is found empty, then it sets $P(\hat{T}=t)=0$, and all other probabilities are uniformly scaled, so they still sum to one. If the agent finds the treasure, it is retired from the simulation. For this simple case the Bayesian model is not strictly necessary, but it will allow us to smoothly integrate later modifications. Here it just prevents the agent from revisiting any empty locations, which is arguably the best possible performance for an agent without any additional information.   

\begin{figure}
\centering
\begin{tikzpicture}[y=18cm, x=.8cm,scale=0.75]
	\draw (0,0) -- coordinate (x axis mid) (10,0);
    	\draw (0,0) -- coordinate (y axis mid) (0,0.22);
    	\foreach \x in {0,...,10}
     		\draw (\x,1pt) -- (\x,-3pt);
     	\foreach \x in {1,...,10}
			\node at (\x-0.5,-8pt) {\x};
    	\foreach \y in {0.00,0.04,0.08,0.12,0.16,0.20}
     		\draw[thin] (10,\y) -- (-3pt,\y) 
     			node[anchor=east] {\y}; 
	\node[below=0.5cm] at (x axis mid) {Location};
	\node[rotate=90, above=1cm] at (y axis mid) {Probability};

\tikzstyle{box}=[fill=grey,thick];	 

\draw[fill=gray] (0.2,0) rectangle (0.5,0.18028);
\draw[fill=gray] (1.2,0) rectangle (1.5,0.09088);
\draw[fill=gray] (2.2,0) rectangle (2.5,0.09069);
\draw[fill=gray] (3.2,0) rectangle (3.5,0.09114);
\draw[fill=gray] (4.2,0) rectangle (4.5,0.09104);
\draw[fill=gray] (5.2,0) rectangle (5.5,0.09063);
\draw[fill=gray] (6.2,0) rectangle (6.5,0.09146);
\draw[fill=gray] (7.2,0) rectangle (7.5,0.09185);
\draw[fill=gray] (8.2,0) rectangle (8.5,0.0907);
\draw[fill=gray] (9.2,0) rectangle (9.5,0.09133);

\end{tikzpicture}
\caption{The probability of observing an agent going to a specific location, if the treasure is located in position 1 and there are 10 locations.}
\label{Fig:TreasureDistribution}
\end{figure}
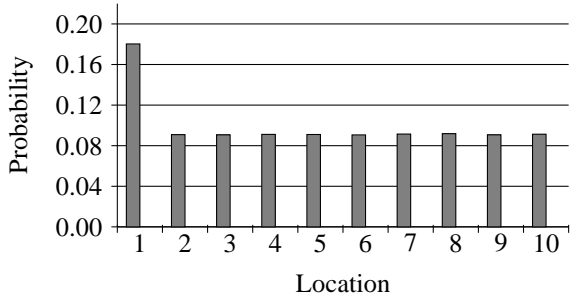
For a world with ten locations it takes on average $\approx$ 5.5 turns to find the location with the treasure. This outperforms an agent which randomly visits (and revisits) locations (10 turns on average to find the treasure), which indicates that the agent is indeed processing information, and subsequently the agent's actions should contain relevant information. Fig.~\ref{Fig:TreasureDistribution} shows the action distribution, gathered by observing 100,000 actions from different non-social agents while the treasure is in position 1. Note that none of the agents act after they found the treasure location, so all observed agents in Fig.~\ref{Fig:TreasureDistribution} are ignorant of where the treasure is, but they know several location where it is not. This is enough processed information to imbue the agent's actions with relevant information.

To model the information another agent would acquire from observing one action from one randomly chosen agent, we assume that all agents are indistinguishable to an observer. If we model their action distribution with a variable called $A$, we can then use that data to compute how much information about $T$, the treasure location, is encoded in $A$. The mutual information in this case computes to $\approx$ 0.042 bit. We can compare those values to the information gained from observing a random location, which is $\approx$ 0.468 bit. So, while inspecting a location contains more information, observing another agent could provide additional information to enhance an agent's performance.

Performance is measured as the ratio of discovered treasure vs. turns. So, if an agent finds treasure on average once every five turns, it then has a performance ratio of $0.2$. The single agent has a performance ratio of $0.180$. This measurement is also identical to the fraction of agent actions that are looking at the right location. This allows us later to evaluate the performance of an agent population, as we do not have to measure the search time, but just measure how many of the agent's actions are going to the treasure location. 

\subsection{Multiple Agent Scenario}

%

The last section indicated that the agent's actions contain relevant information about the treasure location. Therefore, we will now modify the model, so that the agent can integrate data from observing other agents into their internal belief model. 

In the multi-agent model \emph{social} agents will be able to observe the actions taken by other agents, but they will not see the result of this exploration, i.e. know if the visited location is empty. When an agent observes another agent's action $a=A$, it will integrate the obtained information into its own internal model $P(\hat{T})$ by performing a Naive Bayesian Update, based on the statistics for $P(A|T)$ gathered from the non-social statistics in (Fig.~\ref{Fig:TreasureDistribution}). So, its new internal model after observing $a$ is
\begin{equation}
P(\hat{T}|A=a) = \frac{P(A=a|T)}{P(A=a)}P(\hat{T}).
\end{equation}
If an agent finds the treasure, it will be replaced by a new agent, which is simulated by re-initializing an agent's internal model with the uniform distribution.

So, for the multi-agent simulation, all agents start with uniform internal distribution. Each turn the agents then decide their actions, based on their internal model, in the same sequential order. When agents observe other agent's actions, they update their internal model immediately. When agents observe a location, they either update their model if that location is empty, or are replaced by a new agent (have their model reset) if the location contains the treasure.

Note that the Naive Bayesian Update (NBU) works with the assumption that the different sources of information are independently distributed, which is not true in general. NBU still provides good approximations if the dependencies are normally distributed, but in information cascades this is also not the case, as the spread of information through a population is usually self-reinforcing. We still use the NBU, as a more exact Bayesian Update would be nearly impossible to produce, as it would require the agent to remember all previous interactions, and requires statistics on how all other sources of information interact. NBU on the other hand can be done the moment some information becomes available, and the internal belief representation can be represented as a single probability distribution. 

\subsubsection{Single Social Agent}

In the first experiment we examined 10 agents in a world with 10 locations.  All data discussed from here on is the average value for 1,000 simulations, each running for 1,000 turns. Only \textit{one} of the agents has the ability to observe the others. The location of the treasure is fixed, and determined at random at the beginning of the simulation. Unsurprisingly, the remaining non-social agents perform exactly as in the single agent simulation. Their distribution of actions matches the one recorded in Fig. \ref{Fig:TreasureDistribution}.

The social agent in the simulation performs better; reaching a performance of $\approx 0.30$. This agent benefits from the information the other agents gather. As discussed in the ``Digested Information'' argument, the other agents act as information preprocessors for the social agent. Also, note that the distribution of actions of the social agent is even more concentrated on the actual treasure location, hence the mutual information between its actions and the treasure location, $I(A;T) = 0.220$ bits, is higher than the same mutual information for the non-social agents, which was $0.042$ bits.

\subsubsection{All Social Agents}

Given the increase in performance for a single agent, we now assume that the whole population of agents adopts the social update approach, and we examine a simulation where all agents integrate the information gained from other agent's actions.  This turns out to be extremely beneficial. The performance of the overall population, which is also the performance of every separate agent, is $\approx 0.99$. Once the treasure has been located by one agent, all subsequent actions lead to the treasure, and the mutual information between actions and treasure location is nearly maximal, $I(A;T) \approx \log(10)$.

Basically, the relevant information that the treasure is in location $t$ propagates through the agents. It is displayed in an agent's actions, then used to update another agent's internal model. That agent then uses the information to determine which action to take, which is going to be $A = t$. The agent will then find the treasure and reset its internal model. But it will perceive others before it has to act again, biasing its internal model again towards taking action $A=t$. This will continue unless environmental information conflicts with this information, meaning the agent will not find the treasure at the location in which it was looking. In that case, the observed location's probability to contain treasure is set to zero, and the agent will look at other locations. This will initially get the agents to explore all locations until they find the treasure, after which they will all copy each other, finding the treasure every turn from that point onwards. Note, that the treasure does not move when it is found, however the agent who found the treasure resets its internal model (simulating its replacement with a new agent). 

As we see, the important information is preserved by continuously flowing through the agent population. Even when agents retire and are replaced, the information is not lost. This looks like a very desirable feature for an agent population, and therefore the Social Bayesian Update seems like a reasonable adaptation. 

\subsection{Changing World State}

In this section, we will demonstrate how lack of certainty can affect this simulation. We will use the single agent model to motivate the inclusion of noise into our internal Bayesian belief model. 

In the next simulation the locations of the treasure will change during the simulation to different random locations. This will happen every turn with probability of $P(change)=0.01$. On average this should change the location every 100 turns. The behaviour of the agents is left unchanged.

First, let's again take a look at the simulation for a single agent. The performance ratio of the agent drops from $0.18$ for the static world state simulation, to $0.14$ for the simulation where the world state changes. A closer analysis shows that the agent's original behaviour has problems dealing with the new scenario. Consider that the agent visits a location $x$, and finds it empty. Then the probability for $T=x$ will be set to zero in $\hat{T}$. If the location now changes to $T=x$ \textit{after} the agent visited $x$, then the agent will first explore all other locations, finding all of them empty. This, in itself, is not problematic. But once the agent has looked at each locations once, all probabilities are assumed to be zero, given that the agent still assumes there is one, non-moving treasure location. This is inconsistent with the basic properties of probabilities and is a result of the incorrect assumption about the immovability of the treasure location. In this specific implementation the agent now resorts to random search. This behaviour has, as we have seen, a lower performance rate, and therefore lowers the agent's overall performance. 

\subsubsection{Modelling Uncertainty}

To address this problem we can change the internal model to correctly reflect probabilities from the agent's perspective. The treasure changes its location with a probability of $P(change)=0.01$ and relocates to one of the $10$ locations randomly. This can be modelled by assuming that the world is in one of two states. Either, with $P(change)=0.01$, it is in a state where the location has just changed, so $T$ should be uniformly distributed with every $t \in T$  having the probability $P(T=t) = 1 / 10$. The other state, with a probability of $1 - P(change)$, is the one where the treasure location remains unchanged, so the agent should continue to assume the distribution represented by its internal model $\hat{T}$. These two cases can be combined in a weighted sum to determine a new internal distribution $\hat{T'}$. The probability for every state $t$ in this new distribution can be computed as
\begin{equation}
P(\hat{T'}=t) = P(change)\frac{1}{n} + (1-P(change))\cdot P(\hat{T}=t).
\end{equation}
To model the uncertainty, this formula is applied to the agent's internal model each turn after it has completed its action. Note, that this leaves the ordering of probabilities from the most likely to the least likely event intact, unless the probability of change is $1.0$. Therefore, the single agent behaviour with modelled uncertainty performs just as well as the agent without for a non-changing treasure location. But, applying the above uncertainty model to a single agent in a world where the treasure location does change, increases its performance from $0.148$ (for the agent without uncertainty) to $0.180$.

The performance increases because by modelling uncertainty, the agent retains some information about the order in which it explored the previous locations in its internal model. The location that was visited first and found empty subsequently had uncertainty applied to it nine times, once the agent cleared the last, tenth location. It therefore has the largest probability to contain the treasure, and will be the first location to be visited again. This actually reflects the fact that this location is most likely to contain the treasure, since it is unclear when the treasure changed location.


This also shows why modelling the uncertainty works better than simply resetting the probabilities after all locations were visited and found empty. This would reset the internal model and prevent the agent from having to use random search, but it would not preserve the information about the ordering of the previous search, which could be used to the agent's advantage.

\subsection{Uncertainty and Social Bayesian Update}

In this section, we examine a population where all agents model the change uncertainty and also perform the social Bayesian Update. As a result, the agent's performance drops to 0.1, which is equivalent to chance. Closer analysis shows that the whole agent population is always exploring the same location, and the 0.1 average performance is simply the result of the treasure randomly moving to this location from time to time. The agent population here is subject to an information cascade that synchronizes the whole population. But compared to the all-social agent population with internal certainty, the agents cannot reliably check that a certain location is wrong, so after the initial agent breaks the symmetry, the repeated exposure to other agent's social signals will always override their own internal uncertain beliefs. So, while the Social Bayesian Update is beneficial for agents in some cases, it turns out that it can be harmful, specifically when combined with a more accurate model of uncertainty. This is similar to how bounded rationality, i.e. the inability to internally represent certainty, facilitates convergence in social Bayesian network learning \citep{acemoglu2011bayesian}. The difference here is that the lack of internal certainty is not caused by a limitation of the agent, such as cost of internal representation, but motivated by an increase in performance resulting from a more exact modelling of the noise present in the environment.

\subsection{Partial Observability}

One way to reduce the probability for convergence is the reduction of network connectivity \citep{gale2003bayesian}. Currently, the agents live in a neighbourhood of a fully connected graph, being able to observe all other agents. The next simulation has changing treasure locations and an all social, internally uncertain agent population. Unlike the previous models, only a fraction of the other agents' actions can be observed. Every time an agent takes an action, each other agent has a probability of $p_o$ to observe this action and update its internal model. Whether an agent can observe a specific action is determined for each observing agent separately. This creates several simulations interpolating between two previously studied cases. If $p_o = 0$, then the model would be identical to the non-social agent simulation, and if $p_o = 1$, then it would be identical to one in which all agents could observe each other, which leads to a feedback loop and very bad performance ratios. 

\subsubsection{Changing Observation Probability for all Agents}

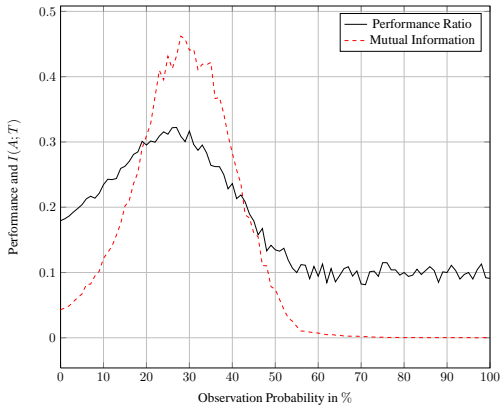
\begin{figure}
\vspace{0.3cm}
\centering
\begin{tikzpicture}[scale=0.5]

\begin{axis}[
xlabel=Observation Probability in $\%$,
ylabel = Performance and $I(A;T)$,
width = 13cm,
xmin = 0,
xmax = 100,
grid = both
]
\addplot[no marks, black] file{PerPerObs.txt};
\addlegendentry{Performance Ratio};
\addplot[no marks, red, dashed] file {MIperPerc.txt};
\addlegendentry{Mutual Information};
\end{axis}
\end{tikzpicture}

\caption{Average performance of an agent population, and the mutual information between its actions and the treasure location, depending on the probability to observe the actions of other agents.}
\label{fig:perObs}
\end{figure}

Varying the parameter $p_o$ for all agents results in performance ratios as depicted in Fig.~\ref{fig:perObs}. As expected, the extremal points are characteristically similar in performance to the non-social and all-social models. In the case where no agents observe each other, the agents find the treasure on average 0.18 times per turn.  The performance ratio increases as the chance to observe other agents increases, up to $\approx 30~\%$ observation probability, where all agents have a performance ratio of $\approx 0.32$. Increasing the observation probability further however, lowers the performance down to approximately 0.1 at an observation probability of $50~\%$ and above.

The second graph (dotted red) in Fig.~\ref{fig:perObs} is the mutual information between the agent's actions $A$, and the treasure location $T$. We see that $I(A;T)$ has the same value as for a non-social agent when the observation probability is zero, it then rises to a peak of $\approx 0.45$ bits for an observation probability of $30~\%$. The mutual information then decreases for larger observation chances, down to zero mutual information for values above $60~\%$.

\subsubsection{Changing Observation Probability for one Agent}

If the observation probability is understood as the result of an agent's effort invested in observing others, then it could be treated as a behavioural parameter that the agent, or at least the process that governs the adaptation of agents, could control. This could be realized by deliberately degrading the agent's sensors to save resources in case of an adaptation process on the agent's population, or by simply discarding some of the sensor input at random if this is realized as an agent strategy. In this context, it would make sense to ask if an individual agent could perform better than the rest of the population by unilaterally changing its probability to observe others.

Given that the actions of the remaining population provide a high degree of mutual information, it might be useful to obtain more of this information than others do. On the other hand, there were indications that taking in too much information from other agents might override the information from the non-agent environment, and thereby degrade the agent's performance. So deliberately lowering the social information intake might also improve the agent's performance compared to the rest of the population.

In the next simulation we will look at one agent that can change its observation probability independently from the rest of the population.  The observation probability for an agent determines how well it can see others, not how well it can be seen. That means that whenever this agent could observe another agent's action, its own observation probability would be used to determine whether this agent could actually sense what action the other agent took.

All other agents in the simulation have a fixed observation probability of 30~\%, since this was the value that led to the best performance for the overall population, and also encoded the most information. 

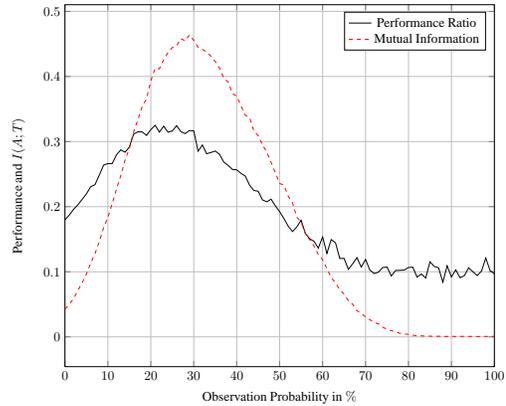
\begin{figure}
\vspace{0.3cm}
\centering
\begin{tikzpicture}[scale=0.5]

\begin{axis}[
xlabel=Observation Probability in $\%$,
ylabel = Performance and $I(A;T)$,
width = 13cm,
xmin = 0,
xmax = 100,
grid = both
]
\addplot[no marks, black] file{PerPerObs30.txt};
\addlegendentry{Performance Ratio};
\addplot[no marks, red, dashed] file {MIperPerc30.txt};
\addlegendentry{Mutual Information};
\end{axis}
\end{tikzpicture}

\caption{Performance of a single agent, and the mutual information between this agent's actions and the treasure location, depending on the probability to observe the other agents in the population. All other agents observe each other with a probability of 30\%.}
\label{fig:per30}
\end{figure}

In Fig.~\ref{fig:per30} we see the resulting performance ratio and mutual information $I(A;T)$ for varying $p_o$ for the one agent that can change its observation probability. Overall, the graph looks very similar to the previous graph in Fig.~\ref{fig:perObs} where all agents could change their observation probability. The performance for one agent is still optimal at $\approx 30\%$. Scaling down the observation probability to zero obviously leads to the same performance as the non-social agent. Increasing observation probability further also results in lowering the performance to approximately 0.1.

This is particularly interesting because, for this specific simulation, it creates something akin to a game theoretic equilibrium at the 30~\% point. All other factors being equal, even if all agents could change their own observation probability at will, none of them could change it away from 30~\% without also decreasing their performance.


\subsection{Relevant Information Analysis}

So far, we have computed the mutual information between the agent's actions and the environment as a measure of how much information their collective actions provide about the state of the environment to an observer. We will now compare this mutual information to the actual relevant information for different performance levels. This will demonstrate that higher observation probabilities are characterized by processing information that is not necessary, indicating the perpetuation of false beliefs in the agent population.


\subsubsection{RI(u) for the Treasure Hunter Model}

The relevant information for the treasure hunter model is determined by the distribution of the treasure, encoded in $T$, and a specific agent's action distribution, encoded in $A$. Both random variables are defined over the same alphabet, which corresponds to all possible locations in the world.

As relevant information is a property of the environment, and not of a specific agent, it therefore considers all possible strategies $p(a|t)$, regardless of how any specific agent would acquire the information needed to actually implement this strategy. To determine the value for $RI(u)$ we have to answer the question, which joint distribution of $A$ and $T$ having at least a performance level of $u$ has the lowest mutual information?


For our specific example of a world with ten locations we can compute the relevant information function as 
\begin{equation}
RI(u) = \log(10) + \left(u \log(u) + (1-u)\log \left(\frac{1-u}{9} \right) \right). 
\label{equ:RIgraph}
\end{equation}

\begin{figure}
\vspace{0.3cm}
\centering
\begin{tikzpicture}[scale=0.5]

\begin{axis}[
xlabel=Performance Ratio,
ylabel = Mutual Information $I(A;T)$,
width = 13cm,
xmin = 0,
xmax = 0.5,
ymin = 0,
grid = both
]
\addplot[black] file{optimalRI.txt};
\addlegendentry{Relevant Information};
\addplot[only marks,mark = x,red,mark size = 2pt] file {30tradeoff.txt};
\addlegendentry{Other Agents 30\% Observation Prob.};
\addplot[only marks,mark =*,blue,mark size = 1pt] file {tradeoff.txt};
\addlegendentry{All Agents change Observation Prob.};
\node [pin=0:0\% observation prob.] at (axis cs: 0.17935 , 0.04284723) {};
\node [pin=100:30\% observation prob.] at (axis cs: 0.300504,	0.45755) {};
\node [pin={[pin distance = 2cm]90:50\% observation prob.}] at (axis cs: 0.1331,	0.110222) {};
\end{axis}
\end{tikzpicture}

\caption{Relevant Information trade-off curve (black line) and points indicating the mutual information and performance for different observation probabilities. 
}
\label{fig:compare}
\end{figure}
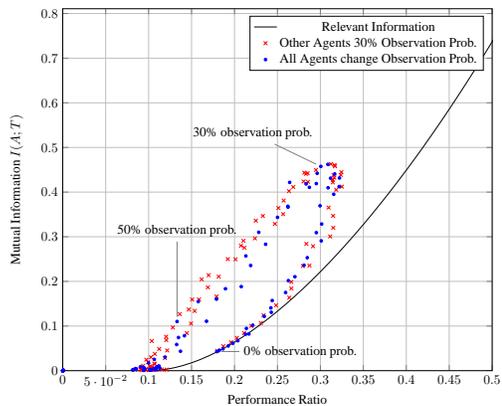

Note that this function computes the minimal mutual information for being on a specific performance level $u$, not for having a strategy that at least has the performance level $u$. However, looking at the actual function, which can be seen in Fig.~\ref{fig:compare}, it becomes clear that the function is, for values of $u$ over 0.1, strictly increasing. Therefore, the minimal mutual information for a specific performance level above 0.1 is also the actual relevant information needed to perform at least that well.
The previous distinction is necessary, because in this case it is necessary to process information to have a performance level lower than 0.1. A performance of 0.1 can be achieved with a random strategy, and therefore has no relevant information. Eq.(\ref{equ:RIgraph}) reflects this, as it is zero for $u = 0.1$. For values of $u$ lower than 0.1 the function in Eq.(\ref{equ:RIgraph}) computes values higher than zero, which would be the information necessary to actually perform \textit{at} this level. One would have to actively avoid the treasure. But by previous definition relevant information should return the information needed to at least attain a specific level, and since random performs better, and has no relevant information, all performance levels below $u = 0.1$ have zero relevant information.


The data points plotted in Fig.~\ref{fig:compare} are taken from the two previous simulations, those where all agents changed their observation probability, and those where only one agent changed its observation probability and all other agents had an observation probability of 30 \%. Each point is the combination of the mutual information $I(A;T)$ and the achieved performance ratio for a specific percentage of observation probability. Different observation probabilities result in different strategies, i.e. different conditional probabilities $P(A|T)$.

The data points gathered here are, as expected, all above or on the RI trade-off curve. The pattern of values are very similar for both simulations. For an observation probability of 0.0 the data point is located at a performance of 0.18, and actually on the trade-off curve. As the observation probability increases, so does the performance. The strategies remain on the trade-off curve at the lower percentages of observation probability, and since the trade-off curve is strictly increasing, so does the encoded relevant information. 

As the observation probability increases we see that the resulting data points leave the trade-off curve, which means the resulting strategies encode more mutual information about the environment than is necessary. The strategies resulting from further increases in observation probability are located in the upper loop where they gravitate towards a point of no mutual information and a performance of 0.1. This indicates that they also encode more information about the environment than necessary. 

Comparison of the mutual information in the actual strategies to the actual relevant information illustrates how observing more and more agents leads to processed information which might not necessarily be relevant. The strategies with low observation probability are located on the relevant information trade-off curve, meaning they are efficient in the sense that they do not process non-relevant information. Those strategies which are subject to the information cascade on the other hand do display a lot of information about the environment in their actions which is non-relevant. At the same time, as seen here, their performance diminishes as well. Fortunately for the agent population, the point where agents display the most relevant information about the environment is also close to the point where the agent performs best, so it would be possible for an agent population, which could adjust their observation probability, to stabilize at a point which benefits all agents the most. 


\section{Conclusion}

Our results indicate that a noisy internal representation seems to be an important factor for the convergence of information cascades, specifically those where the agents perpetuate information that leads to wrong internal beliefs, since the agent cannot, with certainty, reject certain social information. In general, the problem arises in scenarios where signals gained from other agents overpower the agent's private observations and are not as independently generated as the naive Bayesian update models it. On the other hand, the information from other agents is also helpful, and can improve an agent's performance in our model. The interesting observation here is that both things can be influenced by how many other agents an agent randomly observes. Too little, and the agent loses the social information, too much, and the agent population will converge, but possibly on the wrong belief. 

Our relevant information analysis also shows how the quality of the information suffers when more and more agents observe each other. For very low observation probabilities, all the information processed is relevant and agents only display relevant information in their actions. When the agents observe more, their performance gets better still, but we see that they start to pass on information that is incorrect and perpetuate it, sometimes leading to false convergences. As the ``good'' relevant information is still improving, this unnecessary information seems acceptable, but if we increase the observation chance even further, then we see that the performance suffers and the information provided by the agents is mostly wrong. 

In our model however, there exists a point where agents both perform optimally and provide the most information, so a population of agents could adapt to a strategy where they discard a certain percentage of their observations, and perform well. In this case, the agents would basically determine the observation network of the model themselves.
The exact parameter of how many of one's observations one should discard is, of course, model dependent. For example, if the number of agents increases, then it likely takes more observations for total convergence, but a lower observation probability could be sufficient to provide enough social information to overpower the agent's internal beliefs. This is interesting if this is seen as a model for fads and fashions. If an agent, adapted to a population with a specific degree of connectivity, adapts to discard a certain percentage of social information, and is then transplanted to another population, with different parameters, it might become much more susceptible to false self-perpetuating beliefs. The same is true for a population of agents that manages to change their environment in a way that radically changes how much they can observe others.

\section{Acknowledgements}
This research was supported by the European Commission as part of the CORBYS (Cognitive Control Framework for Robotic Systems) project under contract FP7 ICT-270219. The views expressed in this paper are those of the authors, and not necessarily those of the consortium.

\footnotesize
\bibliographystyle{apalike}
\bibliography{alife}

\end{document}